\documentclass[%
reprint,
preprint,
onecolumn,
showpacs,preprintnumbers,
amsmath,amssymb,
aps,
]{revtex4-1}

\usepackage{graphicx}
\usepackage{dcolumn}
\usepackage{bm}

\begin{document}

\preprint{ }

\title{
Pressure-induced unusual metallic state in EuNiO$_3$
}

\author{Hisao Kobayashi}
 \email{kobayash@sci.u-hyogo.ac.jp}
\affiliation{
Graduate School of Material Science
and Center for Novel Material Science under Multi-Extreme Conditions, University of Hyogo, Koto Hyogo 678-1297, JAPAN
}

\author{Shugo Ikeda}%
\affiliation{
Graduate School of Material Science
and Center for Novel Material Science under Multi-Extreme Conditions, University of Hyogo, Koto Hyogo 678-1297, JAPAN
}


\author{Yoshitaka Yoda}
\affiliation{
Japan Synchrotron Radiation Institute, Hyogo 679-5198, JAPAN
}%
\author{Naohisa Hirao}
\affiliation{
Japan Synchrotron Radiation Institute, Hyogo 679-5198, JAPAN
}%
\author{Yasuo Ohishi}
\affiliation{
Japan Synchrotron Radiation Institute, Hyogo 679-5198, JAPAN
}%

\author{J. A. Alonso}
\affiliation{
Instituto de Ciencia de Materiales de Madrid, CSIC, Cantoblanco, E-28049 Madrid, SPAIN
}%
\author{M. J. Martinez-Lope}
\affiliation{
Instituto de Ciencia de Materiales de Madrid, CSIC, Cantoblanco, E-28049 Madrid, SPAIN
}%
\author{R. Lengsdorf}
\affiliation{
II. Physikalisches Institut, Universit\"at zu K\"oln, Z\"ulpicher Stra\ss e 77, 50937 K\"oln, GERMANY
}%
\author{D. I. Khomskii}
\affiliation{
II. Physikalisches Institut, Universit\"at zu K\"oln, Z\"ulpicher Stra\ss e 77, 50937 K\"oln, GERMANY
}%
\author{M. M. Abd-Elmeguid}
\affiliation{
II. Physikalisches Institut, Universit\"at zu K\"oln, Z\"ulpicher Stra\ss e 77, 50937 K\"oln, GERMANY
}%

\date{\today}

\begin{abstract}
The perovskite antiferromagnetic ($T_{\rm N}$ $\sim$ 220 K) insulator EuNiO$_3$ undergoes at ambient pressure 
a metal-to-insulator transition at $T_{\rm MI}$ = 460 K which is associated 
with a simultaneous orthorhombic-to-monoclinic distortion, leading to charge disproportionation. 
We have investigated the change of the structural and magnetic properties of EuNiO$_3$ with pressure (up to $\sim$ 20 GPa) 
across its quantum critical point (QCP) using low-temperature synchrotron angle-resolved x-ray diffraction 
and $^{151}$Eu nuclear forward scattering of synchrotron radiation, respectively. 
With increasing pressure we find that after a small increase of $T_{\rm N}$ ($p$ $\leq$ 2 GPa) 
and the induced magnetic hyperfine field $B_{\rm hf}$ at the $^{151}$Eu nucleus ($p$ $\leq$ 9.7 GPa), 
both $T_{\rm N}$ and $B_{\rm hf}$ are strongly reduced and finally disappear at $p_{\rm c}$ $\cong$ 10.5 GPa, 
indicating a magnetic QCP at $p_{\rm c}$. 
The analysis of the structural parameters up to 10.5 GPa reveals no change of the lattice symmetry 
within the experimental resolution. 
Since the pressure-induced insulator-to-metal transition occurs at $p_{\rm IM}$ $\cong$ 6 GPa, 
this result implies the existence of an antiferromagnetic metallic state between 6 and 10.5 GPa. 
We further show from the analysis of the reported high pressure electrical resistance data on EuNiO$_3$ 
at low-temperatures that in the vicinity of the QCP the system behaves as non-Fermi-liquid, 
with the resistance changing as $T^{\rm n}$, with n=1.6, whereas it becomes a normal Fermi-liquid, 
n = 2, for pressures above $\sim$15 GPa. 
On the basis of the obtained data a magnetic phase diagram in the ($p$, $T$) space is suggested.
\begin{description}
\item[PACS numbers]
71.30.+h, 62.50.-p, 71.28.+d, 76.80.+y
\end{description}
\end{abstract}
\maketitle
\section{\label{sec:Intro}Introduction}

The study of magnetic quantum transitions in strongly correlated electron systems has been 
the subject of continuous interest due to the observation of novel ground states 
near/at the magnetic-to-nonmagnetic transition leading to quantum critical point (QCP). 
Known examples are heavy-fermion metals where non-Fermi-liquid (NFL) phases 
(e.g. CeCu$_{6-x}$Au$_x$ \cite{lohneysen96,lohneysen98} and YbRh$_2$Si$_2$ \cite{custers03}) 
and unconventional superconductivity 
(e.g. CePd$_2$Si$_2$ \cite{mathur98} and UGe$_2$ \cite{saxena00}) appear. 

Another promising but different class of systems for such studies are strongly correlated 
transition metal oxides \cite{khomskii14} (e.g. $RM$O$_3$ perovskites, $R$ = rare earth ion; $M$ = transition metal). 
The key aspect of these materials is that the interplay between spin, charge, 
and orbital degrees of freedom leads to the existence of several competing phases and in turn to complex 
and unusual phase diagrams \cite{imada98,dagotto05}. 
Of particular interest are magnetically ordered transition metal oxides, 
in which a metal-to-insulator (MI) transition as well a magnetic-to-nonmagnetic transition 
can be tuned by external pressure. 
In this context, a central issue is what is the impact of those degrees of freedom 
on the ground state properties when a magnetic insulator is tuned to a nonmagnetic metal across a QCP. 

In this respect, the perovskite rare earth nickelates $R$NiO$_3$ ($R$ $\not=$ La) are excellent candidates 
for such studies as they exhibit an insulating antiferromagnetic (AF) ground state: 
The phase diagram of the $R$NiO$_3$ series is shown in Fig. 1. 
They display a well-defined MI transition at a temperature $T_{\rm MI}$ which increases 
with decreasing the size of $R^{3+}$ ion ($T_{\rm MI}$ = 130 K (Pr), cc, 600 K (Lu) \cite{torrance92,medarde97}). 
Simultaneously the lattice symmetry changes from orthorhombic ($Pbnm$) to monoclinic ($P2_{1}/n$) 
which contains two nonequivalent Ni-sites (NiO$_6$) octahedra with slightly different Ni-O bond lengths, 
indicating charge disproportionation (CD), 2Ni$^{3+}$ $\rightarrow$  Ni$^{3+\delta}$ + Ni$^{3-\delta}$ 
\cite{alonso99_0,alonso99_1,mizokawa00}. 

Another though related interpretation of the phenomena occurring at the MI transition is 
the important role of oxygen holes \cite{mizokawa00,johnston14}, which play a very important role in these materials 
with negative charge-transfer gap \cite{zaanen85}. 
This picture is actually very close to the picture of disproportionation to more and less covalent Ni sites, 
proposed by Goodenough \cite{goodenough96} and in Ref. \cite{zhou04_0}. 
As a matter of fact, in reality both these pictures are different sides for the same phenomenon, 
with the charges on Ni sites more different for smaller rare earths $R$ in the series $R$NiO$_3$, 
and more equivalent for larger $R$. 
Further on we denote both these pictures as CD.

Recently, resonant x-ray diffraction \cite{staub02,scagnoli06}, Raman spectroscopic studies \cite{zaghrioui01} 
and high resolution x-ray absorption at the Ni $K$-edge \cite{medarde09} indicate the existence of CD 
in the whole $R$NiO$_3$ series. 
Moreover, very recent studies on $R$NiO$_3$ using soft x-ray magnetic powder diffraction \cite{bodenthin11} 
demonstrate that the $R$NiO$_3$ compounds have very similar electronic and magnetic states 
despite the large variation of the value of their $T_{\rm MI}$.

At low temperatures, the transition to AF ordered state is also related to the size of the $R^{3+}$ ion, 
i.e. for large $R^{3+}$ ions ($R$ = Pr and Nd) the MI transition occurs simultaneously 
with an antiferromagnetic (AF) ordering of the (Ni) sublattice (i.e. $T_{\rm N}$ $\approx$ $T_{\rm MI}$), 
whereas for smaller $R^{3+}$ ($R$ = Sm $\rightarrow$  Lu) ions,  $T_{\rm N}$ is much lower than $T_{\rm MI}$ 
(e.g. for EuNiO$_3$, $T_{\rm N}$ = 220 K and $T_{\rm MI}$ = 463 K). 
The magnetic structure in the Ni sublattice of $R$NiO$_3$ proposed by neutron-powder diffraction 
\cite{alonso99_0,garcia92,rodriguez98} has a magnetic propagation vector $\bf k$=(1/2,0,1/2) 
and consists of an up-up-down-down stacking of Ni magnetic moments. 
An alternative non-collinear magnetic structure with the same propagation vector is recently suggested 
by resonant soft x-ray magnetic diffraction studies on $R$NiO$_3$ \cite{scagnoli06,bodenthin11,scagnoli08}.

As mentioned above, according to the magnetic phase diagram of the $R$NiO$_3$ series at ambient pressure (see Fig. 1), 
the ground state changes from an antiferromagnetic insulating ($R$ $\not=$ La) 
to a nonmagnetic metallic state ($R$ = La) through a QCP.  
Figure 1 shows the phase diagram in terms of the tolerance factor $t$, which reflects 
the degree of distortion of perovskites and is determined by ratio of 
the relative $R$-O and Ni-O bond lengths, $d_{R- \rm O}$ and $d_{\rm Ni-O}$, 
$t = (d_{R- \rm O})/\sqrt{2} (d_{\rm Ni-O})$. 
As the distortion is larger in for small $R^{3+}$ ions, $t$ increases as the radius $R^{3+}$ increases. 
The figure also shows that $t$ increases with increasing pressure. 
Regarding the effect of external pressure on $T_{\rm MI}$ and $T_{\rm N}$ in $R$NiO$_3$, 
only few compounds have been investigated up to very high pressure until now, 
in particular the pressure dependence of $T_{\rm N}$ in them. 
While the initial change $T_{\rm N}$ ($p$) up to about 2.8 GPa reveals a small increase of $T_{\rm N}$ 
with pressure for $R$ = Sm, Eu and Gd \cite{zhou08}, for NdNiO$_3$ and PrNiO$_3$ ($T_{\rm N}$ = $T_{\rm MI}$) 
$T_{\rm N}$ is strongly reduced with pressure \cite{obradors93,canfield93} 
and even suppressed to zero in PrNiO$_3$ across a QCP \cite{zhou05}. 
In the latter case a broad NFL behavior in the vicinity of the QCP has been reported \cite{zhou05}.

In this work we study the pressure effect on the structural, transport and magnetic properties of 
one of perovskite nickelates, EuNiO$_3$. 
We have selected EuNiO$_3$ ($T_{\rm MI}$ = 463 K and $T_{\rm N}$ = 220 K) which is one of 
typical $R$NiO$_3$ compounds with $T_{\rm MI}$ $>$ $T_{\rm N}$ and thereby allows one to investigate 
the evolution of the magnetic state under high pressure across a QCP. 
The presence of $^{151}$Eu M\"ossbauer isotope in EuNiO$_3$ allows us to realize 
such an investigation at a microscopic level using the $^{151}$Eu nuclear forward scattering (NFS) 
of synchrotron radiation - a technique based on the M\"ossbauer effect. 
The application of the $^{151}$Eu NFS technique allows one to probe the magnetic state of the Ni sublattice 
of EuNiO$_3$ under pressure via the induced magnetic hyperfine (hf) field at the $^{151}$Eu nuclei 
which results from the ordered Ni magnetic moment. 
Some preliminary data which demonstrate the applicability of this technique on EuNiO$_3$ 
have been published by some of the authors elsewhere (\cite{lengsdorf04}, see below). 
The present $^{151}$Eu NFS data on EuNiO$_3$, in combination with low temperature synchrotron angle-resolved 
x-ray diffraction measurements up to about 20 GPa reveal a magnetic QCP at 10.5 GPa. 
Since the high pressure resistance data on EuNiO$_3$ reveal a pressure-induced insulator-to-metal (IM) transition 
at $p_{\rm IM}$ $\sim$ 6 GPa \cite{lengsdorf04}, the results suggest that for pressures 
between 6 and 10.5 GPa the pressure-induced metallic state is magnetically ordered. 
Similar metallic antiferronagnetic state was proposed in Ref. \cite{mazin07}, and recently observed 
in strained multilayers of PrNiO$_3$/PrAlO$_3$ \cite{hepting14}.  
The presence of this state can be possibly explained by the picture of spin density wave \cite{lee11}. 
As the lattice symmetry in this state in our system remains monoclinic, we suppose 
that certain CD exist in this case too, although we do not have definite proof of that. 
We further show from the analysis of the high pressure electrical resistivity data on EuNiO$_3$ \cite{lengsdorf04} 
that in different pressure ranges beyond the QCP both NFL and Fermi liquid (FL) regimes are realized.

It is worthwhile to mention that beside $R$NiO$_3$ bulk samples, rare earths nickelates are very actively 
studies nowadays as thin films and multilayers \cite{frano13,chakhalian14}. 
In particular, one can effectively control their properties by using epitaxial strain 
(and spatial confinement) induced by the substrate, which is rather similar, 
but not identical to the pressure effects studied in this paper.

\section{\label{sec:Exper}Experimental details}

A polycrystalline sample of EuNiO$_3$ was prepared under an oxygen pressure of 200 bars. 
Details of the preparation and characterization were published elsewhere \cite{alonso95}. 
The $^{151}$Eu NFS experiments were carried out under pressure and at low temperature 
using a clamp-type diamond-anvil cell (DAC) on beamline BL09XU at SPring-8. 
The first excited nuclear state of $^{151}$Eu has an energy of 21.541 keV (resonance energy) 
and a half lifetime of 9.7 ns. 
The pulsed synchrotron radiation was monochromatized to a bandwidth of 1.8 meV 
at the resonant excitation energy of $^{151}$Eu nuclei by a high-resolution monochromator. 
The monochromatized x-ray transmitted through the sample was detected with the stacked Si-avalanche photodiodes. 
The storage ring was operated in a special bunch mode where the interval between successive single bunches 
is 165.2 ns which is much longer than the half lifetime of the first excited nuclear state. 
The powder-samples were loaded with ruby chips into a sample cavity of a 0.5 mm diameter in a 0.2 mm thick 
Inconel 625 alloy gasket and mixtures of methanol-ethanol as a pressure-transmitting medium 
to ensure hydrostatic conditions. 
Pressure was calibrated by measuring the wavelength shift of the $R_1$ luminescence line of ruby chips 
in the clamp-type DAC at room temperature.

The x-ray diffraction data were collected under pressure up to approximately 20 GPa at 8 K 
by the angle-dispersive technique and using an image-plate detector on beamline BL10XU at SPring-8. 
The incident x-ray wavelength was 0.4153 \AA, which was calibrated by measuring 
the x-ray diffraction pattern of CeO$_2$ at ambient conditions. 
The powder-samples were loaded into a He-gas driven DAC with ruby chips and He 
as a pressure-transmitting medium to ensure hydrostatic conditions. 
Pressure was calibrated by measuring the wavelength shift of the $R_1$ luminescence line of ruby chips 
in the DAC at 8 K.

\section{\label{sec:R&D}Results and Discussion}
\subsection{\label{sec:EuNFS}High pressure $^{151}$Eu nuclear forward scattering}

As mentioned above, the $^{151}$Eu NFS of synchrotron radiation allows one to probe the magnetic state 
of the Ni sublattice of EuNiO$_3$ under pressure via the induced magnetic hf field $B_{\rm hf}$ 
at the $^{151}$Eu nuclei which results from the ordered Ni magnetic moment. 
$B_{\rm hf}$ originates from the exchange (transferred) field due the admixture of the magnetic ($^7F_1$) excited state 
into the nonmagnetic ($^7F_0$) ground state. 
The magnitude of $B_{\rm hf}$ depends both on the size and relative orientation of the Ni moments 
around the $^{151}$Eu nuclei. 
The first attempt to demonstrate the applicability of this technique on EuNiO$_3$ has been reported 
in Ref. \cite{lengsdorf04}. 
In this work the authors only measured two pressure points (9.5 and 14.4 GPa) 
and detected no magnetic signal at 14.4 GPa, and thus these preliminary measurements provided no information 
about the pressure dependence of $T_{\rm N}$ or $B_{\rm hf}$, 
which is necessary to construct a ($p$, $T$)-magnetic phase diagram. 
In the present work using the high pressure  $^{151}$Eu NFS technique we performed systematic measurements 
of the pressure dependences of $T_{\rm N}$ and $B_{\rm hf}$ of EuNiO$_3$ 
and were able to locate the magnetic QCP and thus to construct the ($p$, $T$)-magnetic phase diagram of EuNiO$_3$.

Figures 2 (a) and (b) show some selected  $^{151}$Eu NFS spectra at different pressures and temperatures, 
both in the paramagnetic stats (7.3 GPa at 300K and 10.7 GPa at 5 K) 
and magnetically ordered state (4.7, 7.3, and 9.7 GPa at 5 K). 
In all spectra, we observe quantum beats due to electric quadruple and/or magnetic hf interactions 
which cause splitting of the nuclear levels and thus lead to a constructive interference of the photons 
emitted from these energy levels. 
As seen in Fig. 2 (a), the frequencies of quantum beats in the $^{151}$Eu NFS spectra 
observed below 9.7 GPa at 5 K are higher than those in the spectrum at 10.7 GPa and 5 K. 
These high frequencies of the quantum beats come from large energy splitting due to magnetic hf splitting 
of the $^{151}$Eu nuclear levels in the magnetically ordered state. 
In comparison, the feature of low frequency in the $^{151}$Eu NFS spectrum on the paramagnetic state 
at 7.3 GPa and 300 K (Fig. 2 (b)) is similar to that at 10.7 GPa and 5 K shown in Fig. 2 (a). 
These results suggest that magnetic ordering in EuNiO$_3$ disappears at 5 K above 10.7 GPa.

The fits to the $^{151}$Eu NFS spectra were performed using the program package MOTIF \cite{shvydko00}, 
applying the full dynamical theory of nuclear resonant scattering and 
including the diagonalization of the complete hyperfine Hamiltonian. 
The spectra at 5 K and 10.7 GPa and at 7.3 GPa and 300 K can be fitted by assuming a pure electric quadrupole interaction, 
indicating a paramagnetic state in EuNiO$_3$ at 5 K and 10.7 GPa.  
However, the $^{151}$Eu NFS spectra in the magnetically ordered state at 5 K below 9.7 GPa 
were impossible to fit by only quadrupole interaction; 
therefore they were fitted by assuming a combined quadrupole and magnetic hf interactions. 
In such a case, we have considered the magnetic structure of EuNiO$_3$ at ambient pressure 
defined by the magnetic propagation vector {\bf k} = (1/2,0,1/2) as determined 
by neutron-powder diffraction \cite{rodriguez98}. 
In this magnetic structure, the single Eu site is subdivided into two magnetically nonequivalent Eu sites 
with the ratio of 1:1. 
One Eu site is sandwiched between two [111] (in cubic setting) layers with parallel spins, 
i.e., surrounded by six Ni atoms with spin up and two Ni atoms with spin down, sees a transferred hf field, 
while the other Eu site is nonmagnetic owing to the cancellation of the Ni antiparallel sublattice 
magnetic moments at these Eu sites \cite{rodriguez98}.

As shown in Fig. 2 (a), the $^{151}$Eu NFS spectra at 5 K below 9.7 GPa were well fitted 
by assuming two different Eu sites with the ratio of 1:1, that is, 
one half of $^{151}$Eu nuclei has $B_{\rm hf}$ with both electric quadrupole and 
magnetic hf interactions, and the other has a small electric quadrupole interaction only. 
Consequently, these results suggest that the magnetic structure in the Ni sublattice of EuNiO$_3$ 
does not change under pressure up to 9.7 GPa. 
The $B_{\rm hf}$ values refined at 5 K are shown in Fig. 2 (a) as a function of pressure. 
As seen in Fig. 3 (a), the refined values of $B_{\rm hf}$ gradually increases with increasing pressure up to 7.3 GPa, 
passes through a maximum around 8 GPa, and then disappears at 10.7 GPa, 
indicating a suppression of Ni magnetic moments.

To obtain the pressure dependence of $T_{\rm N}$, we have analyzed temperature dependences of 
the $^{151}$Eu NFS spectra at different pressures on the basis of the fitting procedure (see above). 
Figure 2 (b) shows that the analytical spectra well reproduce the observed ones within the assumptions given above. 
The values of $T_{\rm N}$ were evaluated from the temperature dependence of $B_{\rm hf}$ 
at different pressures $B_{\rm hf}$($T$, $p$), using the Brillouin function with $S$ = 1/2. 
Figure 3 (b) shows the values of the pressure dependence of $T_{\rm N}$. 
As it is evident from Fig. 2 (b), $T_{\rm N}$ slightly increases with pressure up to about 2 GPa. 
At higher pressures $T_{\rm N}$ decreases and then collapses at about 10.5 GPa, 
indicating the collapse of magnetic order of the Ni moments in EuNiO$_3$. 
The initial increase of $T_{\rm N}$ with pressure ($p$ $\leq$ 2.4 GPa) is in a good agreement 
with the data of EuNiO$_3$ reported from resistivity in Ref. \cite{zhou08} and 
reflects the localized character of the Ni 3$d$-states in the insulating phase of EuNiO$_3$.

Thus, the $^{151}$Eu NFS results reveal that with increasing pressure both $B_{\rm hf}$ 
and $T_{\rm N}$ disappear at $p_{\rm c}$ $\approx$ 10.5 GPa. 
Keeping in mind that EuNiO$_3$ displays the pressure-induced IM transition 
at $p_{\rm IM}$ $\approx$ 6 GPa \cite{lengsdorf04}, 
this result implies that under pressure the ground state of EuNiO$_3$ changes 
from AF insulator to AF metal at $p_{\rm IM}$ and then to a nonmagnetic metal above $p_{\rm c}$. 
The impact of this finding on the complexity ($p$, $T$)-phase diagram of EuNiO$_3$ 
will be discussed in the following sections. 

\subsection{\label{sec:XRD}High pressure synchrotron angle-resolved x-ray diffraction at 8 K}

Figure 4 shows some selected integrated x-ray diffraction patterns of EuNiO$_3$ under pressure at 8 K. 
In the inset of Fig. 4 (a), the diffraction line at $\sim$18 deg. in the pattern corresponds to 
the (224) refraction in the orthorhombic $Pbnm$ structure. 
This diffraction line may split to two (224) and (22$\bar{4}$) refractions in the monoclinic $P2_1/n$ structure. 
But, as it is also known from the structural data on NdNiO$_3$ \cite{staub02} and PrNiO$_3$ \cite{medarde09}, 
the monoclinic distortion in rare earth nickelates with larger rare earths is very small 
and difficult to detect directly. 
That is why it took long time, and required the use of novel, more sophisticated techniques, 
to finally establish that also these systems $R$NiO$_3$, with larger $R$ ions, 
have monoclinic structure at low temperatures \cite{staub02,scagnoli06,medarde09}. 
The situation is the same in our case: as seen in the inset of Fig. 4 (a), we did not observe 
the monoclinic distortion in EuNiO$_3$ within our experimental resolution. 
All diffraction lines in the x-ray diffraction patterns are labeled with the indices of 
the orthorhombic $Pbnm$ structure. 
But these results indicate at least that there is no other pressure-induced structural symmetry 
change up to $\sim$20 GPa at 8 K, within our experimental resolution.

Integrated x-ray diffraction patterns were analyzed with the Rietveld refinement program RIETAN-2000 \cite{izumi00} 
using the orthorhombic $Pbnm$ structure. 
In each x-ray diffraction pattern, the regions at 2$\theta$ $\sim$ 7.2, 8.4 and 12.2 deg. 
were excluded from the refinement procedure where very weak diffraction lines from impurity phases 
were observed. 
All integrated x-ray diffraction lines in the patterns up to $\sim$20 GPa gave good fits 
as shown in Fig. 4. 
It should be noted that these refinement procedures were used to derive individual atomic 
coordination parameters in addition to the lattice parameters.

In Fig. 5 (a), we show the pressure dependence of the refined lattice parameters of EuNiO$_3$ at 8 K. 
As shown in Fig. 5 (a), the pressure dependences of $a$ = $\sqrt{2}a'$, $b$ = $\sqrt{2}b'$, 
and $c$ = 2$c'$ exhibit no discontinuity up to $p$ $\leq$ 10.5 GPa, 
indicating within the experimental resolution no structural phase transition at $p_{\rm IM}$ $\approx$ 6 GPa. 
These pressure variations further reveal that the pressure dependences of $a$, $b$ and $c$ are quite different. 
With increasing pressure, the value of $b$ decreases more rapidly than those of $a$ and $c$ 
and the value of $a$ almost saturates at 10.5 GPa. 
However, no change of the lattice symmetry is observed.

The linear compressibilities $\kappa$ of the lattice parameters were estimated to be 
$\kappa_a$ = 1.11(2)$\times10^{-3}$, $\kappa_b$ = 2.52(3)$\times10^{-3}$, 
and $\kappa_c$ =1.20(1)$\times10^{-3}$ GPa$^{-1}$ below 10.5 GPa. 
The estimated $\kappa$ value of $b$ is twice larger than those of $a$ and $c$. 
Furthermore, the pressure dependences of $a$ and $c$ are different from those at room temperature \cite{lengsdorf04}. 
The bulk moduli $B$ below 10 GPa and above 11 GPa were evaluated based on the Murnaghan equation, 
\begin{equation}
 p = \frac{B}{B'}[(\frac{V_0}{V})^{B'} - 1],
\end{equation}
where $V_0$ is the ambient-pressure volume and $B'$ represents a pressure derivative of $B$. 
The solid lines in Fig. 5 (b) represent the best-fitting curves obtained and the $B$ values were 
refined to be 195(2) and 232(4) GPa below 10 GPa and above 11 GPa, respectively. 
The $B$ value refined below 10 GPa is comparable with those refined at room temperature \cite{lengsdorf04,zhou04}.

Of particular interest is our finding that no change of the lattice symmetry has been observed 
in metallic magnetic ground state for $p_{\rm IM}$ $\cong$ 6 GPa $\leq$ $p$ $\leq$ $p_{\rm c}$ $\cong$ 10.5 GPa. 
One possibility is that the monoclinic distortion associated with CD is too weak 
to be detected by our measurements. 
But the alternative is that the CD is still preserved to some extent also 
in the metallic state in EuNiO$_3$ above 6 GPa. 
This possibility does not contradict our results of $^{151}$Eu NFS in the AF metallic state. 
In this respect, we want to mention that such an unusual ground state has been predicted 
in and observed under high pressure in similar compounds orbitally degenerate compounds (e.g. YNiO$_3$ \cite{mazin07}).

Regarding the observed anomalous pressure dependence of the lattice parameter $a$ 
for $p$ $\geq$ $p_{\rm c}$ $\cong$ 10.5 GPa, it is obvious that the anomaly of $a$ is not related to 
the pressure-induced IM transition at $p_{\rm IM}$ $\cong$ 6 GPa 
but rather corresponds to the transition from the antiferromagnetic metallic 
to a nonmagnetic metallic state at $p_{\rm c}$, i.e. at the magnetic QCP of EuNiO$_3$. 
The origin of this anomaly will be discussed in Section C.

\subsection{\label{sec:C}Magnetic and electronic transitions versus structural parameters}

In the following, we would like discuss the structural response to the pressure-induced IM transition 
and magnetic QCP in EuNiO$_3$. 
In the inset of Fig. 5 (a), we show the pressure dependence of the effective bandwidth $W$ 
as deduced from the structural parameters, which reveals a significant increase 
around the pressure-induced IM transition ($p$ $\geq$$p_{\rm IM}$ $\cong$ 6 GPa). 
$W$ is known to be related to the ligand-to-metal hybridization (effective $t_{pd}$ hopping) 
and can be described in $R$NiO$_3$ compounds in terms of the Ni-O bond length $d_{\rm Ni-O}$ and 
the Ni-O-Ni bond angle $\theta$ by the relation of $W \sim \cos(\pi -\theta)/d_{\rm Ni-O}^{3.5}$ \cite{harrision80}. 
Thus an increase of $W$ implies a corresponding increase of the effective hopping $t_{pd}$ 
which is expected as the system is tuned to a metallic state. 
The averaged $d_{\rm Ni-O}$ and $\theta$ values of EuNiO$_3$ under pressure were evaluated from 
the refined lattice and individual atomic coordination parameters 
to estimate the averaged $W$ value $\langle W \rangle$ under pressure \cite{comm}. 
Obviously the pressure-induced increase of $W$ above the pressure-induced IM transition 
($p$ $\geq$ $p_{\rm IM}$ $\cong$ 6 GPa) reflects the onset of the metallic state in EuNiO$_3$.

On the other hand, $T_{\rm N}$ is also related to $t_{pd}$ ($W$) through the perturbation formula 
for insulators \cite{khomskii14}: 
\begin{equation}
 T_{\rm N} \sim \mathcal{J} \sim t_{dd}^2 \left( \frac{1}{U_{dd}} + \frac{1}{\Delta +\frac{U_{pp}}{2}} \right), 
 \: t_{dd}=t_{pd}^2/\Delta
\end{equation}
where $U_{dd}$ is the on-site $d$-$d$ Coulomb interaction energy, 
$t_{dd}$ the effective $dd$ hoping matrix element, $\Delta$ describes the ligand-to-metal charge-transfer energy, 
and $U_{pp}$ is the Coulomb repulsion of the two holes at the oxygen site. 
As shown in Fig. 2 (b), the pressure dependence of $T_{\rm N}$ in the insulating phase of EuNiO$_3$ 
($p$ $\leq$ $p_{\rm IM}$ $\cong$ 6 GPa) reveals an increase with pressure up to about 2.4 GPa, 
followed by a decrease upon approaching the pressure-induced IM transition, 
whereas $\langle W \rangle$ monotonously increases in the whole insulator phase ($p$ $\leq$$p_{\rm IM}$ $\cong$ 6 GPa). 
Such a deviation from a linear correlation between $T_{\rm N}$ and $W$ in magnetic insulators \cite{zhou08} 
reflects the proximity to a crossover from localized to itinerant electronic behavior in EuNiO$_3$ above about 2.4 GPa.
This is in agreement with the picture proposed by Zhou $et$ $al$. \cite{zhou03} 
to explain the pressure-induced magnetic properties in the $R$NiO$_3$ series.

Now, we would like to provide an explanation of the observed anomalous behavior of the lattice parameter 
across the QCP at $p_{\rm c}$ $\cong$ 10.5 GPa. 
The pressure-induced change of the lattice parameter $a$ is governed by two competing mechanisms 
that act in opposite ways: (i) the decrease of the average Ni-O bond length $\langle d_{\rm Ni-O} \rangle$ 
due to compression of the NiO$_6$ octahedra which leads to a gradual $decrease$ $of$ $a$ with pressure; 
and (ii) the increase of the average Ni-O-Ni bonding angle $\langle \theta \rangle$ 
which results in a reduction of the tilting of NiO$_6$ octahedra and thereby an $increase$ $of$ $a$ with pressure. 
We argue that below $p_{\rm c}$ $\cong$ 10.5 GPa the compression of $\langle d_{\rm Ni-O} \rangle$ 
(i) is the dominating factor and prevails in the pressure dependence of $a$, 
leading to a decrease of $a$ with pressure. 
However, at and beyond the QCP the compression of $\langle d_{\rm Ni-O} \rangle$ becomes weaker 
and comparable to that due to the increase of $\langle \theta \rangle$ (ii). 
We believe that this could be related to a corresponding change of the magnetoelastic coupling 
at and beyond the QCP. 
As a result, the two competing mechanisms become almost equal and compensate each other, 
causing a nearly pressure independent change of the lattice parameter $a$ for $p$ $\geq$ $p_{\rm c}$ $\cong$ 10.5 GPa. 
In such a case, one would anticipate melting of the charge disproportionation in the nonmagnetic metallic state, 
caused by the increased bandwidth.

\subsection{\label{sec:phase}Suggested ($p$, $T$)-phase diagram of EuNiO$_3$}

On the basis of our $^{151}$Eu NFS and low temperature synchrotron x-ray data, 
we would like to propose the ($p$, $T$)-phase diagram for EuNiO$_3$, shown in Fig. 6. 
The overall features of the phase diagram can be summarized as follows: 
at ambient pressure below $T_{\rm MI}$ $\sim$ 460 K the system undergoes a transition 
from the orthorhombic metallic state to a monoclinic CD insulating state. 
The insulating CD state exhibits antiferromagnetic ordering below $T_{\rm N}$ $\sim$ 220 K. 
Under pressure, this AF insulating and CD ground state changes to an AF metallic and CD state 
at $p_{\rm IM}$ $\cong$ 6 GPa. 
At higher pressures the system displays a transition to a nonmagnetic state 
at $p_{\rm c}$ $\cong$ 10.5 GPa i.e. QCP. 
By further increasing pressure beyond the QCP, the system reveals NFL and FL behavior for 
10.5 GPa $\leq$ $p$ $\leq$ 14.8 GPa and 15.9 GPa $\leq$ $p$ $\leq$ 17.5 GPa, respectively. 
The identification of NFL and FL regions are based on the analysis of the original 
low temperature electrical resistance data on EuNiO$_3$ reported in Ref. \cite{lengsdorf04}. 
To analyze the electrical resistance data $R_{\rm abs}$($T$), we used the power-law fitting to 
$R_{\rm abs}$($p$, $T$) of EuNiO$_3$ in the temperature range between 10 and 45 K 
with $\Delta R_{\rm abs}$($T$)(=$R_{\rm abs}$($T$)-$R_0$) $\sim$ $T^{\rm n}$, where n = 2 and 1 $<$ n $<$ 2 for FL 
and NFL behavior, respectively. 
As shown in Figs. 7 (a) and (b), the power-law with n = 1.6 fits the experimental data fairly well 
for $p$ = 11.5, 13.2 and 14.8 GPa, indicating NFL behavior, whereas n = 2 is the best fitting 
for 15.9 and 17.5 GPa, corresponding to FL behavior. 
Our finding of a NFL behavior in EuNiO$_3$ with n = 1.6 is similar to that reported 
from high pressure resistivity data on PrNiO$_3$ \cite{zhou05}. 
The authors show in addition that the suppression of the insulating state of PrNiO$_3$ 
($T_{\rm N}$ = $T_{\rm MI}$ $\sim$ 130K ) above 1.3 GPa is accompanied 
by a transformation to a NFL phase in which the resistivity varies proportional 
to $T^{\rm n}$ with n = 1.33 and 1.6 over a broad pressure range. 
In EuNiO$_3$ with $T_{\rm N}$ $\ll$ $T_{\rm MI}$, we only observe a NFL behavior with n=1.6.

\section{\label{sec:Summary}Summary}

We have investigated the pressure effect up to about 20 GPa on the structural and magnetic properties 
of the antiferromagnetic insulator rare earth nickelate EuNiO$_3$ 
using low-temperature synchrotron angle-resolved x-ray diffraction and $^{151}$Eu nuclear forward scattering (NFS) 
of synchrotron radiation, respectively. 
The $^{151}$Eu NFS technique allows one to probe the magnetic state of the Ni sublattice of EuNiO$_3$ 
under pressure via the induced magnetic hyperfine field at the $^{151}$Eu nuclei which originates 
from the ordered Ni magnetic moments, and thus to investigate the evolution of the magnetic state 
under high pressure across a quantum critical point. 
The experimental results can be summarized as follows.

EuNiO$_3$ shows two transitions: an insulator-to-metal transition 
at $p_{\rm IM}$ $\cong$ 6 GPa (already reported) and magnetic-to-nonmagnetic transition 
with a quantum critical point at $p_{\rm c}$ $\cong$ 10.5 GPa. 
In this context, we would to refer to the recent observation of similar metallic antiferromagnetic phase 
in PrNiO$_3$ strained multilayers \cite{hepting14}.

The analysis of the pressure dependence of the structural parameters revealed a significant increase 
of the effective bandwidth $W$, which is related to effective $t_{pd}$ hopping, 
around the pressure-induced IM transition ($p$ $\geq$ 6 GPa). 
However, we did not detect, within the resolution of the x-ray measurements, 
any anomalies in the lattice parameters at the IM transition at $p_{\rm IM}$ $\cong$ 6 GPa, 
and have seen only slight change of the bulk modulus at the quantum phase transition at $p_{\rm c}$ $\cong$ 10.5 GPa. 
This lets us suggest that most probably the charge disproportionation, existing in $R$NiO$_3$ 
in the insulating phase, survives to certain extent also in the metallic phase. 
Furthermore, we have shown from the analysis of reported high pressure resistance data on EuNiO$_3$ 
at low-temperatures that in the vicinity of the QCP the system behaves as non-Fermi-liquid, 
the resistance behaving as $T^{\rm n}$, with n=1.6, 
whereas it becomes a normal Fermi-liquid, n = 2, for pressures above $\sim$15 GPa. 
Based on all obtained data we propose the ($p$, $T$)-phase diagram for EuNiO$_3$, shown in Fig. 6. 
We feel that the properties of other nickelates of this class might be similar to those revealed here, 
i.e. they may be representative also for other perovskite nickelates.

\begin{acknowledgments}
The $^{151}$Eu NFS and x-ray diffraction experiments under pressure were performed 
at SPring-8 with the approval of the Japan Synchrotron Radiation Research Institute (JASRI)
 (Proposal Nos. 2010A1517, 2008B1460, and 2007A1450). 
M.M.A. and D.K. would like to thank the Deutsche Forschungsgemeinschaft (DFG) for the support through SFB 608. 
The work of D.K. was supported by the German project FOR 1346 and by Cologne University 
within the German Excellence Initiative.
\end{acknowledgments}
%

\newpage
\begin{figure}[tp]
\begin{center}
\includegraphics[width=15cm]{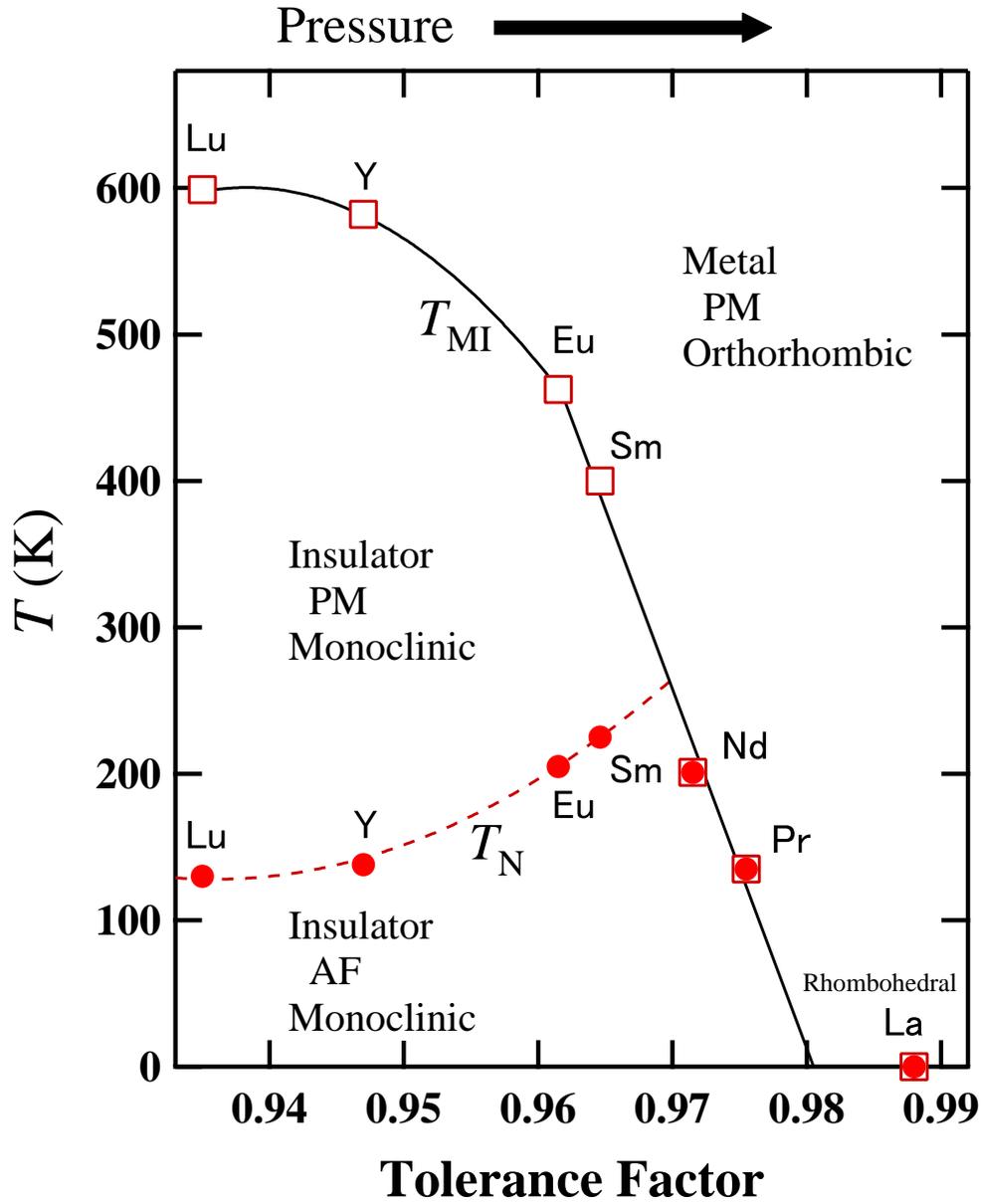}
\end{center}
\caption{
(Color online) The phase diagram of the $R$NiO$_3$ series as a function of the tolerance factor $t$ (see text) 
and external pressure, adapted from data in Refs. \cite{torrance92,medarde97,alonso99_1}. 
PM and AF stand for paramagnetic and antiferromagnetic, respectively.
}
\label{fig:fig0}
\end{figure}

\newpage
\begin{figure}[tp]
\begin{center}
\includegraphics[width=15cm]{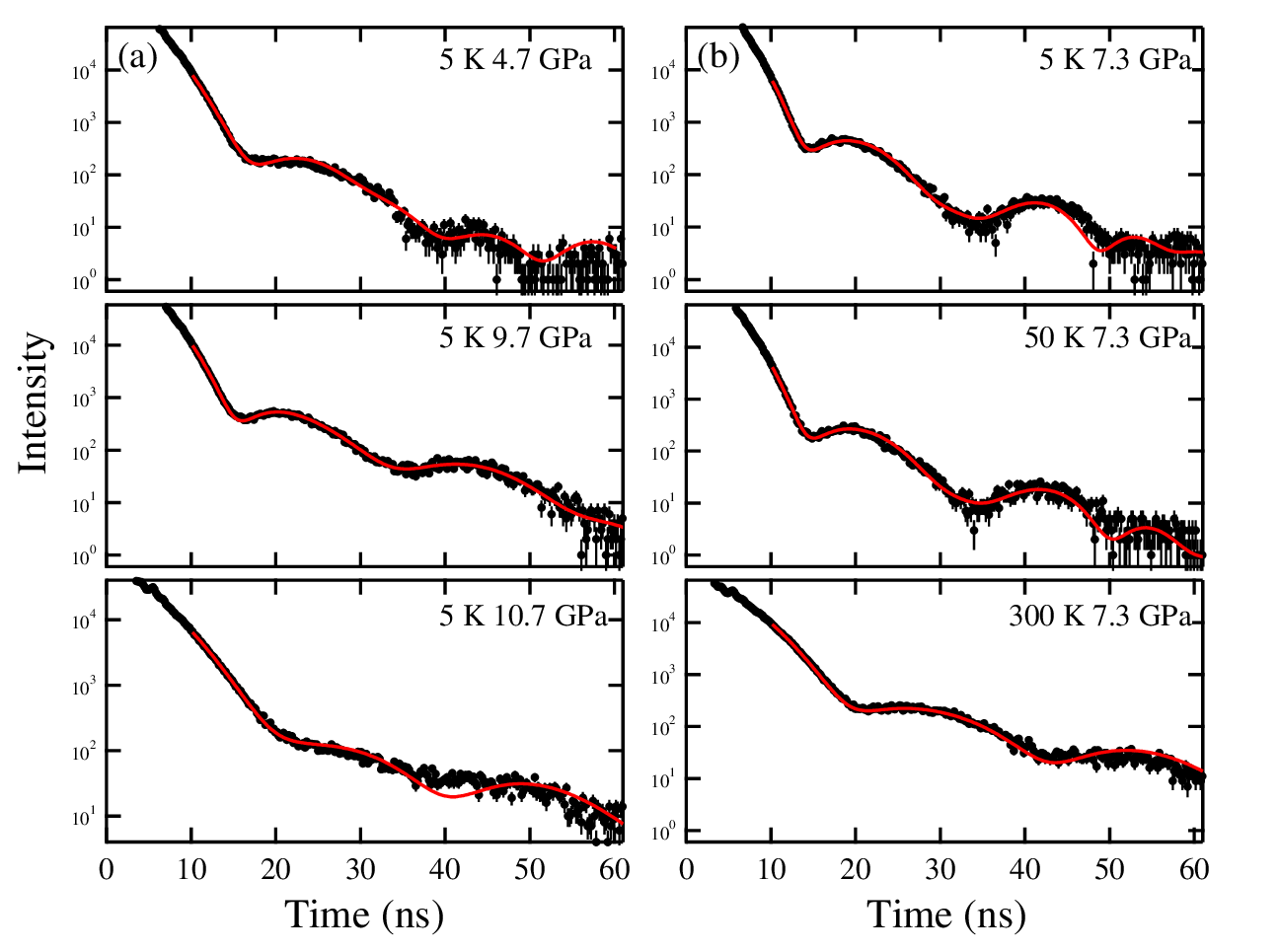}
\end{center}
\caption{
(Color online) Selected $^{151}$Eu nuclear forward scattering spectra of EuNiO$_3$ 
(a) at 5 K under pressures and (b) at 7.3 GPa as a function of temperature. 
The closed circles with error bars indicate the observed spectra and the red solid lines represent 
the best fitted curves obtained using MOTIF \cite{shvydko00}.
}
\label{fig:fig1}
\end{figure}

\newpage
\begin{figure}[tp]
\begin{center}
\includegraphics[width=10cm]{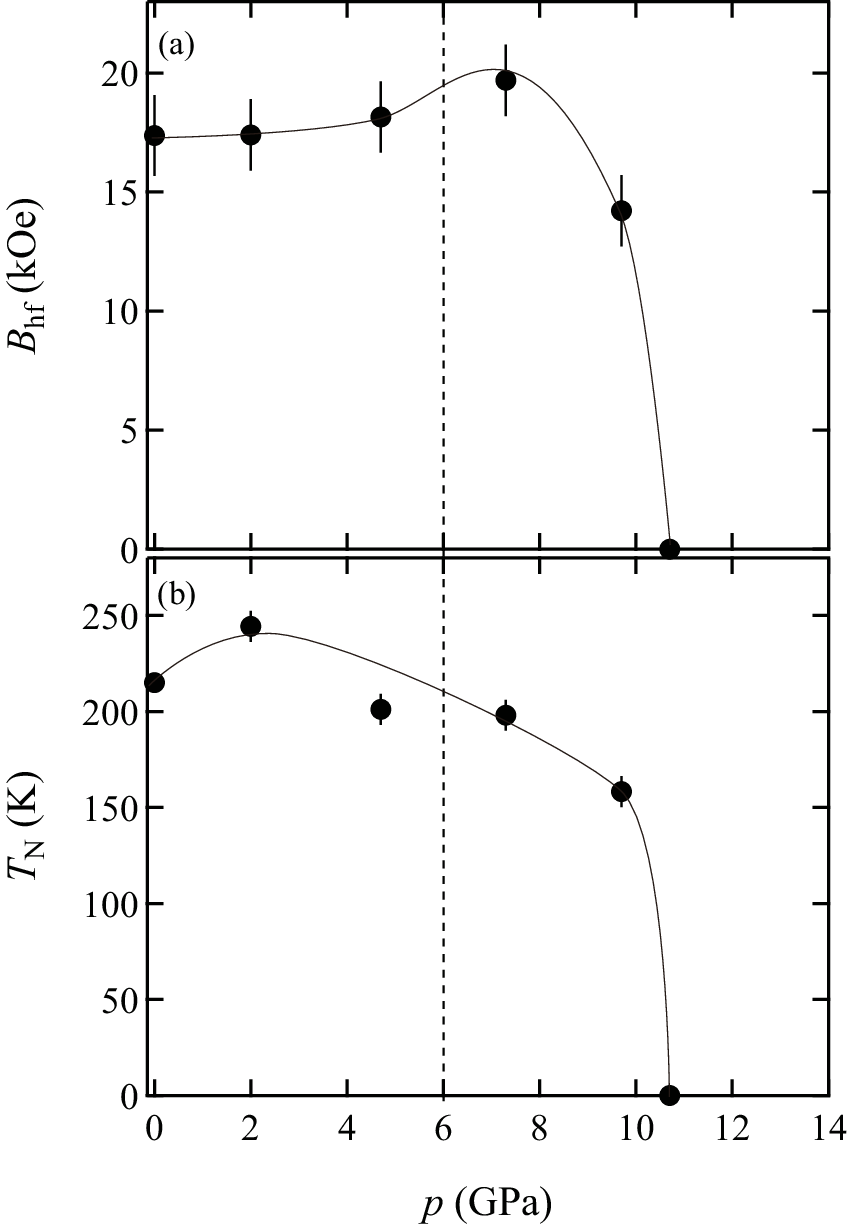}
\end{center}
\caption{
(a) Refined magnetic hyperfine field $B_{\rm hf}$ at 5 K 
and (b) evaluated N\'eel temperature $T_{\rm N}$ of EuNiO$_3$ as a function of pressure. 
At ambient pressure, $B_{\rm hf}$ was refined from the data at 3 K in Ref. \cite{lengsdorf04} 
using MOTIF \cite{shvydko00} by assuming two different Eu sites with the ratio of 1:1. 
Lines through the data points are only a guide to the eye. 
The broken lines in (a) and (b) indicate the pressure-induced insulator-to-metal transition 
at $p_{\rm IM}$ $\cong$ 6 GPa which is taken from the high pressure low temperature resistivity data 
reported in Ref. \cite{lengsdorf04}.
}
\label{fig:fig2}
\end{figure}

\newpage
\begin{figure}[tp]
\begin{center}
\includegraphics[width=12cm]{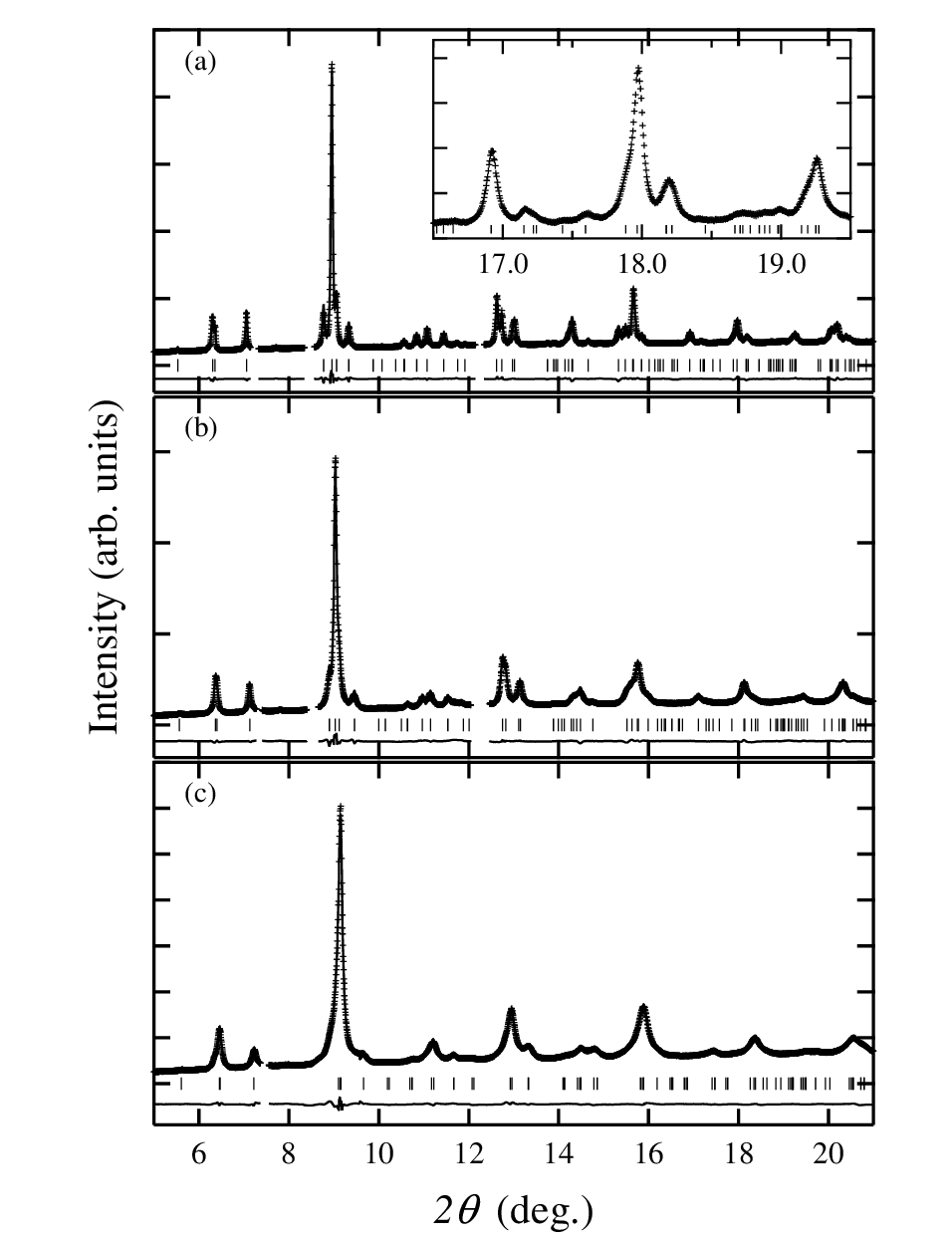}
\end{center}
\caption{
Selected integrated x-ray diffraction patterns of EuNiO$_3$ under (a) 2.0, (b) 7.6, and (c) 17.4 GPa at 8 K, 
where the crosses show the integrated x-ray diffraction intensities. 
The solid lines represent the results of Rietveld refinement fitting \cite{izumi00} 
and the tick marks show the positions of all reflections allowed by the orthorhombic $Pbnm$ symmetry. 
The differences between the integrated and calculated intensities are shown below the tick marks. 
The inset in (a) shows an enlarged high-angle 2$\theta$ region in the pattern at 2.0 GPa, 
where the effect of monoclinic distortion can be expected.
}
\label{fig:fig3}
\end{figure}

\newpage
\begin{figure}[tp]
\begin{center}
\includegraphics[width=12cm]{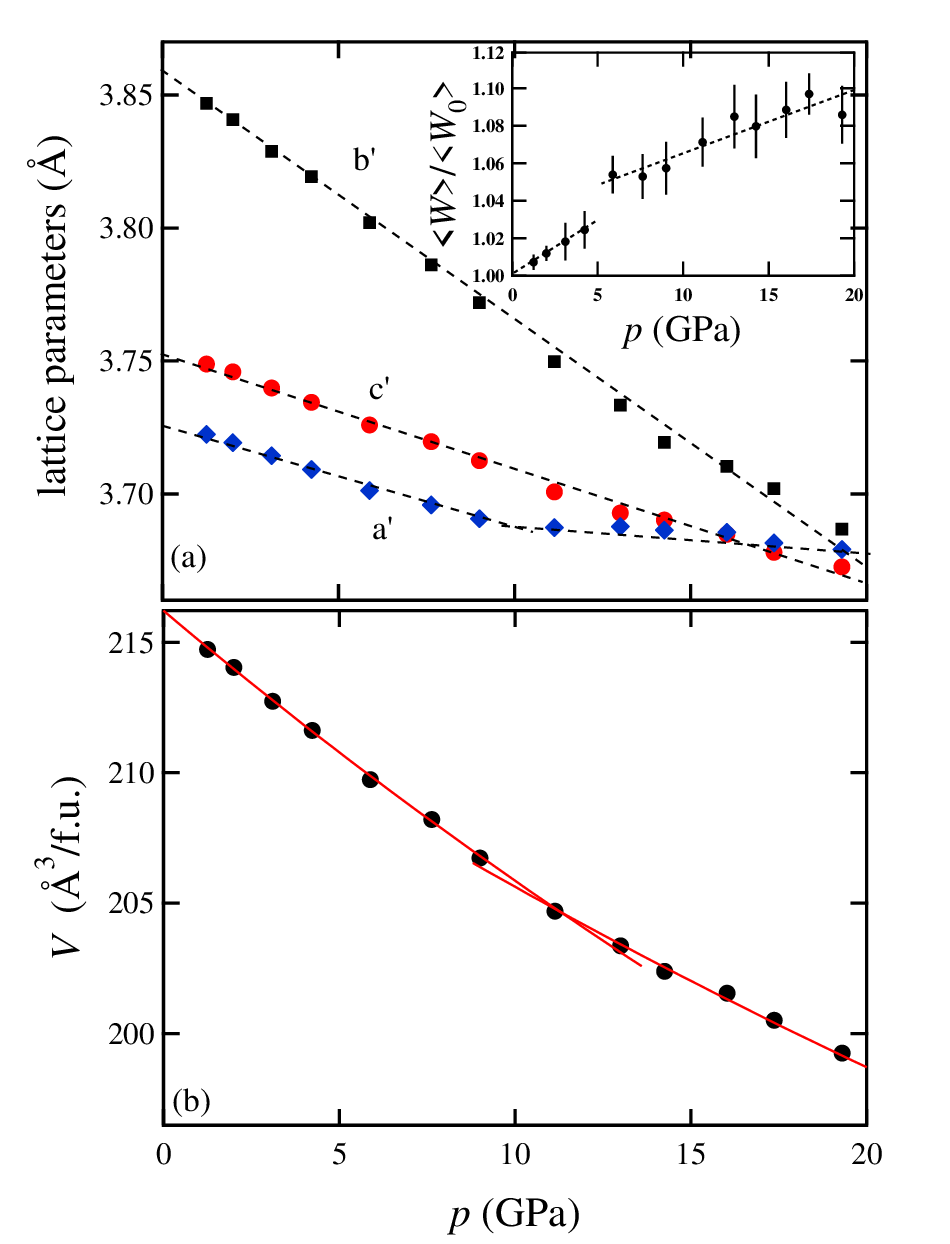}
\end{center}
\caption{
(Color online) (a) Refined lattice parameters, $a$, $b$, and $c$, 
($a$ = $\sqrt{2}a'$, $b$ = $\sqrt{2}b'$, and $c$ = 2$c'$) and (b) evaluated volume, $V$, of EuNiO$_3$ at 8 K 
as a function of pressure. 
The broken lines in (a) are visual guides. 
The solid lines in (b) represent the fitting results based on the Murnaghan equation. 
The inset in (a) shows the pressure dependence of normalized bandwidth $\langle W \rangle$/$\langle W_0 \rangle$ 
which was evaluated by the refined lattice and individual atomic coordination parameters under pressure. 
The broken lines in the inset are visual guides.
}
\label{fig:fig4}
\end{figure}

\newpage
\begin{figure}[tp]
\begin{center}
\includegraphics[width=12cm]{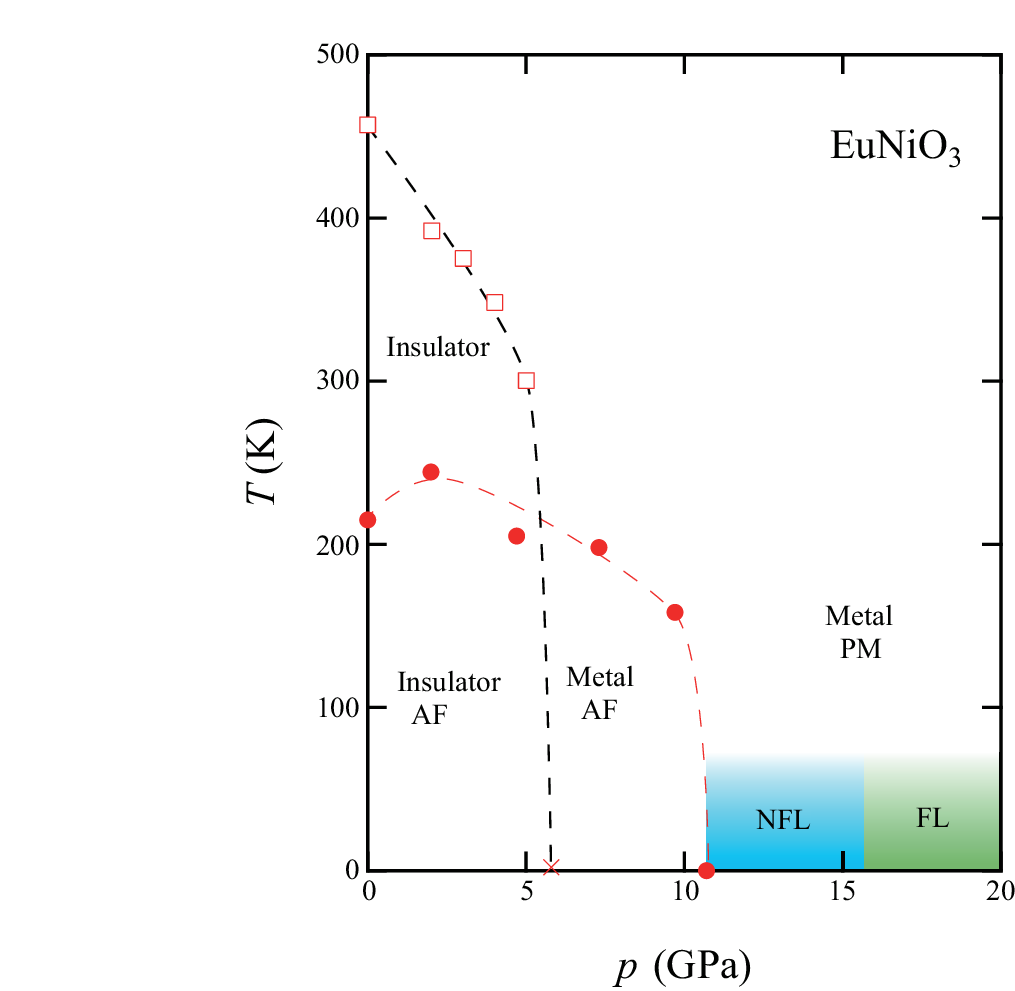}
\end{center}
\caption{
(Color online) Pressure vs. temperature phase diagram of EuNiO$_3$. 
The closed red circles represent N\'eel temperature $T_{\rm N}$ evaluated 
by present $^{151}$Eu nuclear forward scattering data. 
The open squares indicate the pressure dependence metal-to-insulator transition temperature $T_{\rm IM}$ 
reproduced from Ref.  \cite{cheng10}; extrapolated to the critical pressure 
(cross point at low temperatures) indicates the pressure-induced insulator-to-metal transition 
at $p_{\rm IM}$ $\cong$ 6 GPa which is taken from the high pressure low temperature resistivity data 
reported in Ref. \cite{lengsdorf04}. 
Non Fermi-liquid (NFL) and Fermi-liquid (FL) regions above 10.5 GPa are identified 
from the analysis of the original low temperature electrical resistance data on EuNiO$_3$ reported 
in Ref. \cite{lengsdorf04}.
}
\label{fig:fig5}
\end{figure}

\newpage
\begin{figure}[tp]
\begin{center}
\includegraphics[width=15cm]{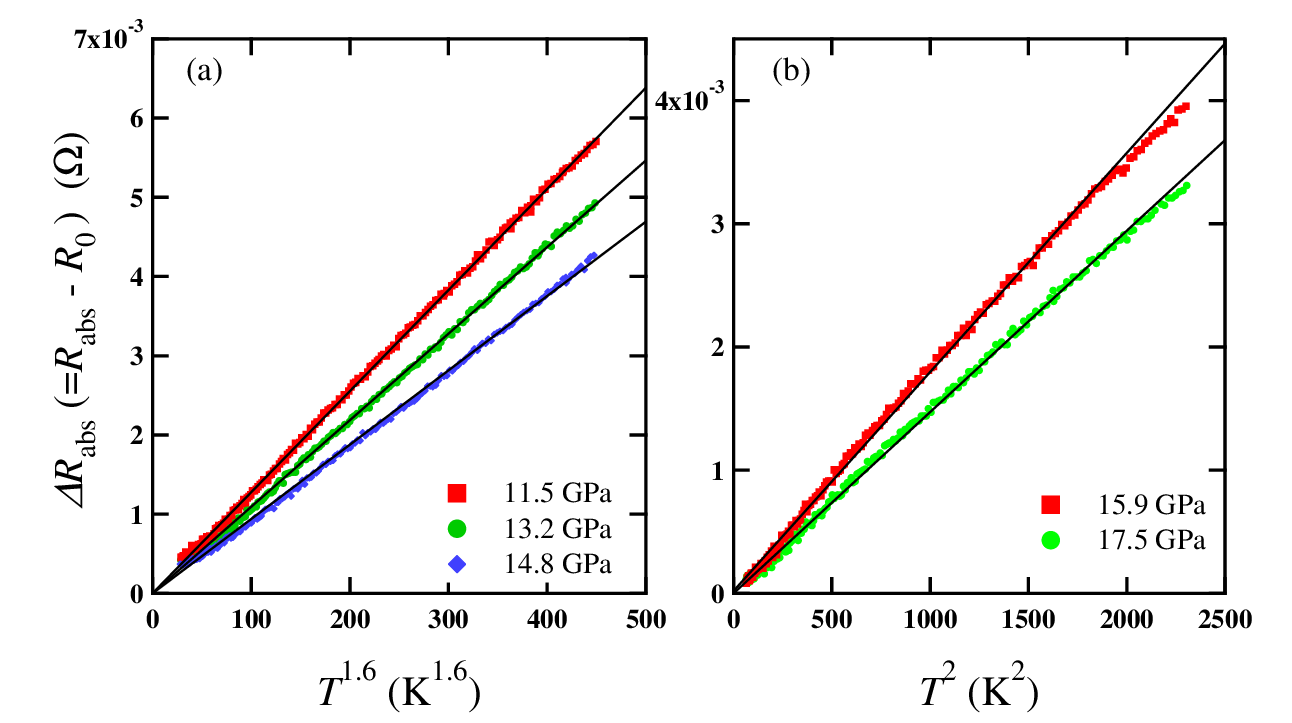}
\end{center}
\caption{
 (Color online) Power-law temperature dependences of electrical resistance $R_{\rm abs}$ 
in the pressure ranges from 11.5 to 14.8 GPa (a) and from 15.9 to 17.5 GPa (b). 
The closed symbols indicate the experimental date taken from Ref. \cite{lengsdorf04}. 
The solid lines represent the fitting results.
}
\label{fig:fig6}
\end{figure}


\begin{thebibliography}{34}
\bibitem{comm}Owing to the two crystallographically nonequivalent O sites in $R$NiO$_3$ 
  with orthorhombic $Pbnm$ structure, there are two different $d_{\rm Ni-O}$ values for the NiO$_6$ octahedron 
  and two different $\theta$ values between the NiO$_6$ octahedra. 
  The averaged $d_{\rm Ni-O}$ and $\theta$ values of EuNiO$_3$ under pressure were evaluated 
  from the refined lattice and individual atomic coordination parameters to estimate 
  the averaged $W$ value $\langle W \rangle$ under pressure.
\end{thebibliography}
\end{document}